\documentclass[12pt]{article}
\pdfoutput=1
\usepackage{relsize}
\usepackage{xspace}
\usepackage{amstext}
\usepackage{graphicx}
\usepackage{calc}
\usepackage{lineno}
\usepackage{hyperref}
\usepackage[all]{hypcap}

\def\beq{\begin{equation}}
\def\eeq#1{\label{#1}\end{equation}}
\def\eeqn{\end{equation}}

\def\beqa{\begin{eqnarray}}
\def\eeqa#1{\label{#1}\end{eqnarray}}
\def\eeqan{\end{eqnarray}}

\let\bar=\overbar

\def\Dslash{\not{\hbox{\kern-4pt $D$}}}
\def\dslash{\not{\hbox{\kern-2pt $\del$}}}

\def\BR{\mbox{\rm BR}}

\def\msb{{\bar{\ssstyle M \kern -1pt S}}}

\ifdefined\bibtexflag
\fi

\makeatletter
\DeclareRobustCommand\bfseries{%
  \not@math@alphabet\bfseries\mathbf
  \fontseries\bfdefault\selectfont\boldmath}

\makeatother

\newcommand{\EE}[1]{\ensuremath{{\cdot}10^{#1}}}

\newcommand{\invfb}{\ensuremath{\mbox{\,fb}^{-1}}\xspace}

\newcommand{\gev}{\ensuremath{\,\text{Ge\kern -0.1em V}}\xspace}
\newcommand{\gevc}{\ensuremath{\,\text{Ge\kern -0.1em V}\!/c}\xspace}
\newcommand{\gevcc}{\ensuremath{\,\text{Ge\kern -0.1em V}\!/c^2}\xspace}
\newcommand{\mev}{\ensuremath{\,\text{Me\kern -0.1em V}}\xspace}
\newcommand{\mevc}{\ensuremath{\,\text{Me\kern -0.1em V}\!/c}\xspace}
\newcommand{\mevcc}{\ensuremath{\,\text{Me\kern -0.1em V}\!/c^2}\xspace}

\newcommand{\babar}{\mbox{%
    \slshape B\kern-0.1em{\smaller A}\kern-0.1em
    B\kern-0.1em{\smaller A\kern-0.2em R}}\xspace}

\newcommand{\epem}      {\ensuremath{e^+e^-}\xspace}

\newcommand{\ellell}    {\ensuremath{\ell^+ \ell^-}\xspace}

\newcommand{\qqbar}{\ensuremath{q\overline q}\xspace}

\newcommand{\ccbar}{\ensuremath{c\overline c}\xspace}

\newcommand{\Bbar}   {\kern 0.18em\overline{\kern -0.18em B}{}\xspace}

\newcommand{\Bu}     {\ensuremath{B^+}\xspace}
\newcommand{\Bub}    {\ensuremath{B^-}\xspace}

\newcommand{\BpBm}   {\ensuremath{\Bu {\kern -0.16em \Bub}}\xspace}
\newcommand{\Bz}     {\ensuremath{B^0}\xspace}
\newcommand{\Bzb}    {\ensuremath{\Bbar^0}\xspace}
\newcommand{\BzBzb}  {\ensuremath{\Bz {\kern -0.16em \Bzb}}\xspace}

\newcommand{\gaga}{\ensuremath{\gamma\gamma}\xspace}

\renewcommand{\BR}{{\cal B}\xspace}

\newcommand{\pepii}{\mbox{PEP-II}\xspace}

\def\Y#1S{\ensuremath{\Upsilon{(#1S)}}\xspace}

{\(\displaystyle}
{\)}

\newcommand{\dtoll}{\ensuremath{D^0\to\ell^+\ell^-}\xspace}
\newcommand{\dtopipi}{\ensuremath{D^0\to\pi^+\pi^-}\xspace}
\newcommand{\dtokpi}{\ensuremath{D^0\to K^-\pi^+}\xspace}
\newcommand{\dtokspi}{\ensuremath{D^0 \rightarrow K_S^0 \pi^0}\xspace}
\newcommand{\dtoee}{\ensuremath{D^0\to e^+e^-}\xspace}
\newcommand{\dtomumu}{\ensuremath{D^0\to \mu^+\mu^-}\xspace}
\newcommand{\dtoemu}{\ensuremath{D^0\to e^\pm\mu^\mp}\xspace}
\newcommand{\dtogamgam}{\ensuremath{D^0\to\gamma\gamma}\xspace}
\newcommand{\dtopizpiz}{\ensuremath{D^0\to\pi^0\pi^0}\xspace}
\newcommand{\dzero}{\ensuremath{D^0}\xspace}
\newcommand{\mdzero}{\ensuremath{m(D^0)}\xspace}
\newcommand{\deltam}{\ensuremath{\Delta m}\xspace}

\newcommand{\pipi} {\ensuremath{\pi^+\pi^-}\xspace}
\newcommand{\piz} {\ensuremath{\pi^0}\xspace}
\newcommand{\pip} {\ensuremath{\pi^+}\xspace}
\newcommand{\pim} {\ensuremath{\pi^-}\xspace}
\newcommand{\kpi}  {\ensuremath{K^-\pi^+}\xspace}

\newcommand{\dStar}{\ensuremath{D^{*+}}\xspace}

\newcommand{\gammasigeff}{6.1\% \xspace}

\newcommand{\gammayield}{-6 \pm 15}

\newcommand{\gammagammasensitivity}{2.4 \cdot 10^{-6}}

\newcommand{\pizsigeff}{15.2\% \xspace}

\newcommand{\kspizsigeffforgammacuts}{7.6 \%}
\newcommand{\kspizsigeffforpizcuts}{12.0\%}

\def\Title#1{\begin{center} {\Large {\bf #1} } \end{center}}

\begin{document}
\ifdefined\linenoflag\linenumbers\fi

\Title{Search for $D^0 \to \ell^+\ell^-$ and $D^0 \to \gamma\gamma$ \\
  at \babar and BESIII}

\bigskip\bigskip


\begin{raggedright}
{\it Alberto Lusiani\index{Lusiani, A.} (on behalf of the \babar and BESIII collaborations)\\
Scuola Normale Superiore e INFN di Pisa\\
\vspace{1ex}
Proceedings of CKM 2012, the 7th International Workshop on the CKM
Unitarity Triangle, University of Cincinnati, USA, 28 September - 2 October
2012}
\bigskip\bigskip
\end{raggedright}

\section{Introduction}

Flavour changing neutral currents (FCNC) leading to the decay of charmed $D$
mesons into \ellell and \gaga are very suppressed in the Standard Model (SM),
and in fact the Glashow-Iliopoulos-Maiani (GIM) mechanism
\cite{Glashow:1970gm} is more effective here than for $B$ and $K$ mesons
because the down-type quark mass differences are relatively smaller than
the up-type ones.
In the SM, the \dtomumu and \dtoee branching fractions are dominated by
long-distance contributions and estimated to be $\BR(\dtomumu) \sim
10^{-13}$ and $\BR(\dtoee) \sim 10^{-23}$~\cite{Burdman:2001tf}, while
\dtoemu violates lepton flavour conservation and is forbidden in the SM. If
neutrino mixing is accomodated in the SM, \dtoemu would anyway happen only
at undetectable small rates.
The SM predictions are all well below the present experimental
sensitivities, but searches for these decays can probe NP contributions
that may provide enhancements of several orders of
magnitude~\cite{Golowich:2009ii}. Furthermore, NP models can affect \dtoll
and $D^0 - \bar{D^0}$ mixing in a correlated way~\cite{Golowich:2009ii}.
The short-distance SM contribution to \dtogamgam is $\simeq
3\EE{-11}$, the dominant long-distance contribution is $\simeq
3.5\EE{-8}$~\cite{Burdman:2003rs}. NP models can enhance the SM rate by
more than 2 orders of magnitude~\cite{Prelovsek:2000xy}.

\section{\babar search for \dtoll}

\babar has searched for \dtoll~\cite{Lees:2012jt} using approximately 468~fb$^{-1}$ of
$e^+e^-$ collisions produced by the \pepii asymmetric-energy $e^+e^-$
collider at and near the $\Upsilon(4S)$ resonance.

\dzero candidates are found by combining pairs of oppositely charged
leptons. Decays to \pipi and \kpi are used as control samples.
$D$ mesons from direct $e^+e^- \to c \bar c$
production are selected by requiring that the \dzero candidate momentum be
larger than 2.4\,\gevc, which is about the kinematic limit for $B \to D^*\pi$,
$D^{*+}\to D^0\pi^+$. Combinatorial background is suppressed by requiring that the \dzero
candidate originate from the decay $D^*(2010)^+ \to \dzero
\pi^+$~\footnote{Here and in the following charge conjugation is implied
unless otherwise stated}.
A fit of the $D^{*+} \to \dzero \pi^+; D^0 \to \ell^+\ell^-$ decay chain is
performed where the \dzero tracks are constrained to come from a common
vertex and the \dzero and slow pion are constrained to form a common vertex
within the beam interaction region.
The $\chi^2$ probabilities of the \dzero and $D^*$ vertices from this fit must
be at least $1$\%.
The reconstructed \dzero mass \mdzero must be within
$[1.65, 2.05]$\,\gevcc and  the mass difference $\Delta m$ must be within
$[0.141, 0.149]$\,\gevcc.  The \dzero and $\Delta m$ mass resolutions, measured
in the \dtopipi sample, are 8.1\,\mevcc and 0.2\,\mevcc, respectively.

Background candidates are either random
combinations of two leptons (combinatorial background), or
\dtopipi decays where both pions pass the lepton identification
criteria (peaking background).
The \dtopipi background is most important for the \dtomumu channel.
Combinatorial background from semileptonic $B$ decays is suppressed by using
a Fisher discriminant~\cite{Fisher:1936} of five variables.
Events with hard initial state radiation that converts to eletron pairs are
suppressed by requiring at least 5 tracks for the \dtoee channel and
at least 4 tracks for the \dtomumu and \dtoemu channels, by
accepting at most 3 electron candidates and by vetoing particles compatible
with being results of photon conversions.
The selection criteria for each signal channel were chosen
to give the lowest expected signal branching fraction upper limit
for the null hypothesis (a true branching fraction of zero)
using the MC samples.

Control samples and sidebands are used to estimate backgrounds.
Combinatorial background is estimated using the upper sideband in the
\dzero mass distribution.  Peaking background is estimated by scaling the
observed \dtopipi yield with the misidentification rates estimated from the
\dtokpi data control sample.

For all three channels, no statistically significant yield is observed
compared with the expected background: 1 event observed vs.\ $1.0 \pm 0.5$
expected background for \dtoee, 2 observed vs.\ $1.4 \pm 0.3$ expected for
\dtoemu and 8 observed vs.\ $3.9 \pm 0.6$ expected for \dtomumu. The
probability of observing 8 events when $3.9 \pm 0.6$ events are expected is
5.4\%.
Accounting also for systematic uncertainties, the 90\% confidence level
(CL) upper limits for the branching fractions are $\BR(\dtoee) <
1.7\EE{-7}$, $\BR(\dtoemu) < 3.3\EE{-7}$ and $\BR(\dtomumu) <
2.6\EE{-7}$~\cite{Lees:2012jt}.

\section{\babar search for \dtogamgam}

The \babar collaboration has conducted a search for
\dtogamgam~\cite{Lees:2011qz} on
470.5\,\invfb of data collected by the $\babar$ detector at the SLAC PEP-II
$e^+e^-$ asymmetric-energy collider operating at $\epem$ CM energies at and
around the $\Upsilon(4S)$ peak. The important
background decay mode \dtopizpiz and the decay mode $\dtokspi$
are also selected and studied in the same analysis. The last channel is
used as reference normalization mode to avoid uncertainties
in the number of $\dStar$.

\dzero candidates are made with pairs of photons and pairs of \piz mesons,
and the resulting \dzero is required to come from the decay $D^*(2010)^+
\to \dzero \pi^+$ ($\dStar$ tag).
The candidates are required to have a momentum $>2.85 \gevc$ and their
invariant mass must be between $1.7$ and $2.1\gevcc$ for the $\dtogamgam$
analysis and between $1.65$ and $2.05\gevcc$ for the $\dtopizpiz$
analysis. The photon candidates are selected to have CM energies between
$0.74$ and $4\gev$, the $\piz$ candidates must have energy larger than
$0.6\gev$.
The \dzero candidates for the \dtokspi reference mode are formed by
combining a $\piz$ candidate as defined above with a $K_S^0$ candidate
consistent with the decay $K_S^0 \rightarrow \pi^+\pi^-$, with a decay
length significance (flight length divided by its estimated uncertainty)
larger than 3.
The \dtogamgam selection rejects events whose photons are compatible with a
\piz decay.
A kinematic fit is applied to the events, requiring the candidate $\dzero$
invariant mass to be between $1.6$ and $2.1\gevcc$. Both the $\dzero$ and
$\pip$ are constrained to originate from a common vertex within the
beamspot to satisfy the $\dStar$ tag requirement.
The $\dtogamgam$ analysis signal efficiency is $\gammasigeff$ with the
corresponding reference mode ($\dtokspi, K_S^0 \rightarrow \pip \pim$)
efficiency at $\kspizsigeffforgammacuts$. The $\dtopizpiz$ analysis signal
and reference mode efficiencies are $\pizsigeff$ and
$\kspizsigeffforpizcuts$, respectively.

\begin{figure}[tb]
\centering
\includegraphics[width=0.5\linewidth]{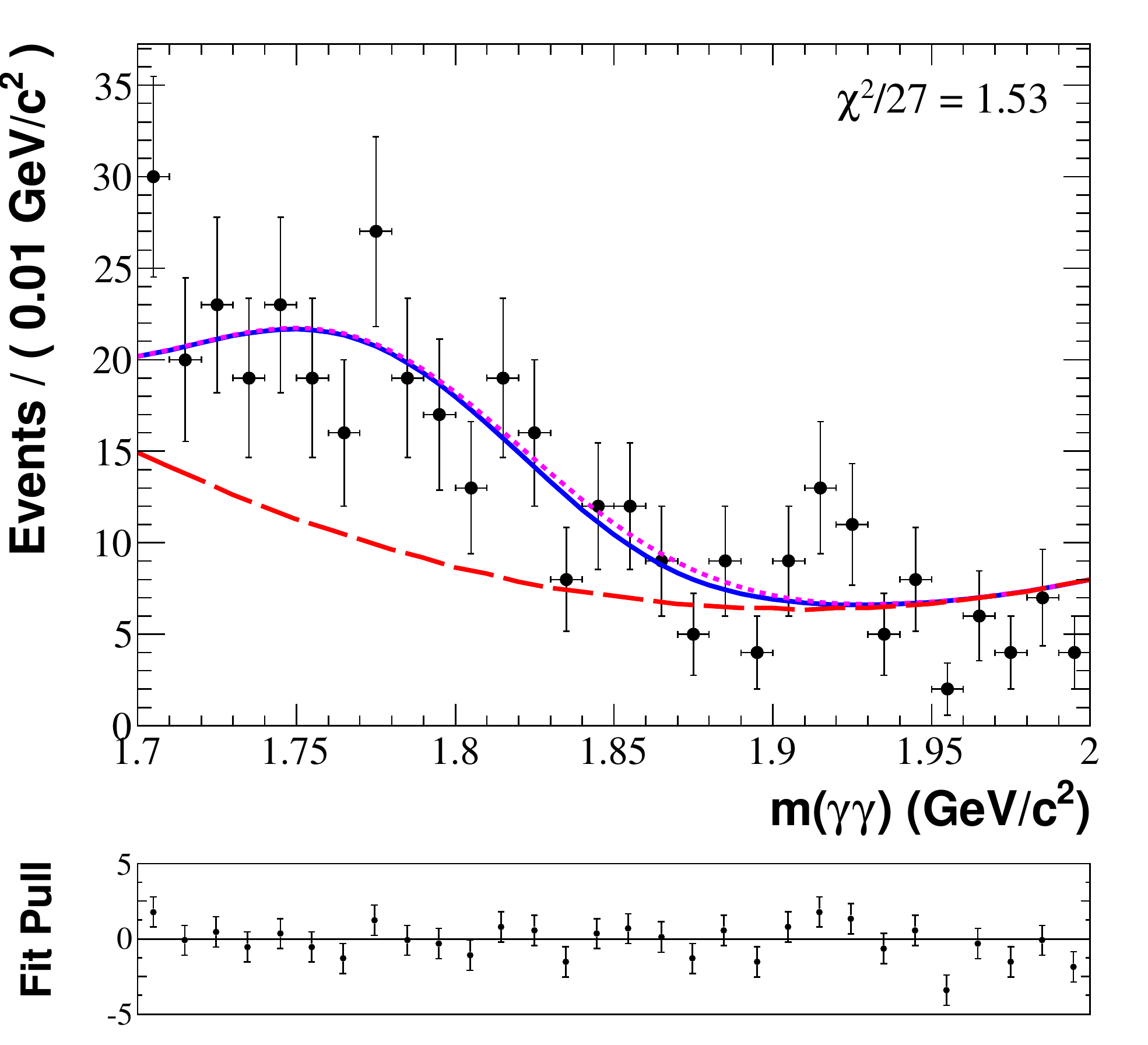}%
\caption{The $\gamma\gamma$ mass distribution for $\dtogamgam$ candidates
  in data (data points). The solid blue curve
  corresponds to the signal plus background fit, where the signal component
  has a slight negative yield, the
  long-dash red curve corresponds to combinatorial background component, and
  the small-dash pink curve corresponds to the combinatoric background plus
  $\dtopizpiz$ background shape. The $\chi^2$ value is determined from
  binned data. The pull
  distribution shows differences between the data and the solid blue curve
  with values and errors normalized.}
\label{fig:gammafit_data}%
\end{figure}

For all modes, signal yields are determined using unbinned maximum
likelihood fits to the invariant mass distribution of $D^{0}$
candidates. The probability distribution functions (PDFs) for signal and
backgrounds have their shapes fixed from Monte Carlo simulation and their
relative normalization fit on data.
The invariant $\gamma\gamma$ mass distribution of the $\dtogamgam$
candidates is shown in Fig.~\ref{fig:gammafit_data} together with the fit
of the signal and background component contributions.
The signal yield is $\gammayield$, consistent
with no $\dtogamgam$ events. Using the $\dtokspi$ yield, $\BR(\dtogamgam)
= (-0.49 \pm 1.23 \pm 0.02)\EE{-6}$  where the
errors are the statistical uncertainty and the uncertainty in the reference
mode branching fraction, respectively.

Several systematic uncertainties cancel partially or
completely by normalizing the yield to the reference mode; the largest
estimated systematic contributions come from uncertainties in tracking
efficiency (1.4\%), \piz veto (1.8\%), photon reconstruction (0.6\%). The
\piz veto performance has been studied on data using the physically
forbidden decay $D^{0} \rightarrow K_{S}^{0} \gamma$.
Accounting for systematic uncertainties, the 90\% CL upper limit for
\BR(\dtogamgam) is determined to be $<
\gammagammasensitivity$~\cite{Lees:2011qz}.

\section{BESIII search for \dtogamgam}

The BESIII collaboration has reported preliminary results on a search for
$D^0 \rightarrow \gamma\gamma$~\cite{Muramatsu:2012nh} based on an
integrated luminosity of ${\sim}2.9$ fb$^{-1}$ of $e^+e^-$ collisions
produced at the BEPCII energy-symmetric ring at $\sqrt{s}=3.773$\,\gev and
recorded with the BESIII detector~\cite{besiii:2010}. The analysis includes
also the measurement of the major background $D^0\to\pi^0\pi^0$.

To select the $D^0\to\pi^0\pi^0$ channel, photons are first paired to form \piz
candidates, which are then re-fitted with a mass constraint. \piz
candidates are combined to form \dzero candidates, whose energy is required
to match half the event energy with a discrepancy $-60\,\mev < \Delta E <
30\,\mev$, where $\Delta E = E_{\pi^0\pi^0} - E_{\text{beam}}$. The signal yield is
obtained with a maximum-likelihood fit to the beam-constrained mass
$M_{bc}$ $\equiv$ $\sqrt{E^2_{\text{beam}} - p_D^2c^2}$ where $E_{\text{beam}}$
is the beam energy and $p_D$ is the $D^0(\to\pi^0\pi^0)$ candidate
momentum. The fit yields $4081\pm117$ signal events.

The selection of $D^0\to\gamma\gamma$ candidates starts with pairing the
two most energetic photons. Events with photons compatible with
$\pi^0\to\gamma\gamma$ are rejected. Radiative Bhabha events are suppressed
by requiring that photons be isolated at least by $20^\circ$ from all tracks
and by requiring that the most energetic track fail the electron
identification.

Finally, the signal yield is extracted with a fit to the $\Delta E$
distribution after requiring that
$1860<M_{bc}<1870$~\mev$/c^2$. Here, the signal shape is fixed from Monte
Carlo. The fit has $\chi^2$ of 63.7 for 76 degrees of freedom and
determines a signal yield of $-2.9\pm7.1$.

No significant signal is found and 90\% CL upper limits are set to
$\BR(D^0\to\gamma\gamma)/\BR(D^0\to\pi^0\pi^0)<5.8\EE{-3}$ and
$\BR(D^0\to\gamma\gamma)<4.7\EE{-6}$~(preliminary)~\cite{Muramatsu:2012nh}.

\section{Conclusions}

There is no evidence of New Physics in the searches that \babar and BESIII have
done for the rare charm decays \dtoll and \dtogamgam. The experimental
sensitivities are still far from the SM predictions, leaving space for
further significant improvements.

\ifdefined\bibtexflag
\bibliography{%
  bibtex/pub-c-95-99,%
  bibtex/pub-d-00-04,%
  bibtex/pub-e-05-09,%
  bibtex/pub-2010,%
  bibtex/pub-2011,%
  bibtex/belle-2003-2005,%
  bibtex/belle-2006-2008,%
  bibtex/tau-lepton,%
  bibtex/pub-extra,%
  bibtex/misc,%
  lusiani-ckm12-procs-bib%
}
\else
\providecommand{\href}[2]{#2}

\fi

\end{document}